\documentclass[endfloats,12pt,prb,preprint,showpacs]{revtex4}

\usepackage{amsmath}
\usepackage{graphicx}

\begin{document}

\date{\today }
\title{Extra energy coupling through subwavelength hole arrays via stochastic resonance}
\author{Jean-Baptiste Masson}
\email{jean-baptiste.masson@polytechnique.edu}
\author{Guilhem Gallot}
\email{guilhem.gallot@polytechnique.edu}
\affiliation{Laboratoire d'Optique et Biosciences, Ecole Polytechnique, CNRS, INSERM, 91128 Palaiseau, France}

\begin{abstract}
\end{abstract}


\maketitle

Interaction between metal surface waves and periodic geometry of subwavelength structures is at the core of the recent but crucial renewal of interest in plasmonics \cite{TK18,TK58,TK93,TK56}. One of the most intriguing points is the observation of abnormal strong transmission through these periodic structures, which can exceed by orders of magnitude the classical transmission given by the filling factor of the plate \cite{TK93}. The actual paradigm is that this abnormal transmission arises from the periodicity, and then that such high transmission should disappear in random geometries. Here, we show that extra energy can be coupled through the subwavelength structure by adding a controlled quantity of noise to the position of the apertures. This result can be modelled in the statistical framework of stochastic resonance. The evolution of the coupled energy with respect to noise gives access to the extra energy coupled at the surface of the subwavelength array.

We investigated the transmission of subwavelength arrays with a controlled quantity of noise added to the position of the subwavelength apertures, in order to study how plasmonic properties would disappear with addition of noise. Most surprisingly, our experimental results demonstrate a strong increase of coupled energy through the array by adding a small quantity of noise to the lattice. This unexpected extra energy coupling could not be modelled using standard plasmonic theories \cite{TK54,TK94,TK74,TK85}, such as Bethe, Bloch waves or CDEW models, since they intrinsically predict maximum energy coupling with perfectly periodic organization. We explain our results using stochastic resonance theory \cite{TP24}, which describes how a system can be optimized by the presence of a certain amount of noise. Furthermore, the evolution of the coupled energy for relatively large values of noise gives access to fundamental characteristics of the establishment of the electromagnetic propagation and coupling at the surface. It allows the direct measurement of the extra energy coupled at the surface of the screen originating from the creation of the abnormal high transmission process.
 
Experiments were made in the terahertz domain since this domain is particularly suited for this study thanks to the lack of plasma resonance on metals in this part of the electromagnetic spectrum, and to the high precision achievable in the screen design.
We designed noised arrays of squared subwavelength apertures by adding Gaussian noise displacements to the hole centres with a standard deviation $\sigma$ (see Fig.~\ref{array}). No deviation ($\sigma=0$) corresponds to perfectly periodic structure, while high deviations deal with fully random aperture positions. Numerous plates have been investigated, one for each noise deviation. Three typical experimental spectra are shown in Fig.~\ref{spectre}. For perfectly periodic arrays ($\sigma=0$), the spectrum exhibits the usual strong asymmetrical first resonance and a second resonance scaling as $\sqrt{2}$. For large deviations ($\sigma=60\pm 0.5\,\mu$m), spectra show a less intense, much wider and symmetrical resonance, and no clear shapes for higher frequencies. For low quantity of noise, a much unexpected behaviour is observed for a small amount of noise ($\sigma=15.4\pm0.5\,\mu$m). Here, the spectrum exhibits a much stronger resonance. This resonance is still asymmetrical, and the complete spectrum exhibits all usual Surface Plasmon Polaritons (SPPs) features, but more energy is coupled through the array. This result is clearly unexpected, and seems to indicate that noise added to the structure increases the coupled energy. To explain these results, the evolution of transmitted spectrum with the amount of noise has been studied. To reduce the influence of spectrum artefacts due to specific patterns generated on the noised arrays, we introduce the transmission of the array $T=E_t/E_0$, where $E_t$ and $E_0$ are the total transmitted and incident energies, respectively. On Fig.~\ref{experience_theorie}, evolutions of $T$ versus $\sigma$ are shown for both arrays of period $L=$300$\,\mu$m (Fig.~\ref{experience_theorie}A) and $L=$400$\,\mu$m (Fig.~\ref{experience_theorie}B). We observe that $T$ increases up to a specific value of deviation, $\sigma_{M}$,  and then decreases. There is a clear resonance of $T$ with respect to noise. For $L=$300$\,\mu$m and $L=$400$\,\mu$m arrays, this relative increase of energy is respectively of $18.2\pm0.01\%$ and $11\pm0.01\%$. The maxima of $T$ correspond to $\sigma_{M}=10.4\pm0.5\,\mu$m and $\sigma_{M}=15.4\pm0.5\,\mu$m, only 3.4\% and 3.8\% of the array period, respectively. At the resonance frequency, the energy loss has been reduced by 2.7 for $L=$400$\,\mu$m (Fig.~\ref{spectre}). For $\sigma \geq \sigma_{M}$, $T$ continuously decreases as one would have previously expected.
 
In a first attempt to describe this unexpected extra energy coupling, several models currently used to describe abnormal transmission through periodic subwavelength arrays have been tested: Block wave theory \cite{LivreKittel}; calculation based on pure diffraction effects, derived on Bethe diffraction theory \cite{TK1}; Composite Diffracted Evanescent Wave theory \cite{TK141,TK140} which provides some extra adjustable parameters; Fano model \cite{TK72}. Results are shown in addition to experimental data in Fig.~\ref{experience_theorie}B. None of them could reproduce the observed features, since they intrinsically rely on the periodicity to explain the resonance frequencies and coupling yield.
   
The question of modelling this unexpected energy coupling will now be addressed. Physical systems where the increase of noise leads to a rise of the output energy can be integrated into the Stochastic Resonance (SR) framework. SR was first introduced by Benzi \textit{et al} \cite{TP23} in order to describe global climate oscillations between ice and temperate ages, showing that white noise could help a bistable system to change state at the same frequency than a very weak periodic coupling interaction. This model was more rigorously treated by McNamara \textit{et al} \cite{TP24}, using master equation formalism in a system of two coupled levels in interaction with a Gaussian noise and showing that specific values of noise could increase the exchange rate of particles between the two levels. Today, SR theories have been extended to non negligible or non periodic coupling, coloured noise, non Markovian processes and finally multilevel systems, and are widely used to model neuronal signal generation, chemical reactions, some colloidal solutions properties, ionic channels activation, public opinion changes, and some complex auto organization processes \cite{TP33,TP29,TP30,TP31,TP32,TP34}.
  
A SR McNamara-like model will be used to describe our experimental data. Among the theories describing the abnormal transmission through subwavelength hole arrays, the Fano model (Figure~\ref{schema}A) draws an interesting parallel between the scattering process through a continuum of states interacting with an isolated state and the diffusion of light through the array \cite{TK72,TK79,TK96}. The continuum of states describes the light directly crossing the array, either through the aperture, or either re emitted by the edge of the holes. And the resonant states are the surface waves. This model allows describing both propagation and scattering on complex arrays and gives good experimental data fits, and can easily be related to SR theories by the existence of discrete levels. Furthermore, it has been observed that the SR gets enhanced if an array of similar elements collectively responds to the same signal. This phenomenon has been termed the array enhanced stochastic resonance (AESR) \cite{TP36,TP37}. We then modeled the present experiment giving the evolution of $T$ versus $\sigma$ as the sum of many individual stochastic resonance (Figure~\ref{schema}B), and $T$ is the result of SPPs scattering with a global stochastic resonance, which is the sum of all SR effects between couples of levels. As a direct consequence of the choice of this model, the fitting function is given by a Lorentzian function \cite{TP24}
\begin{equation}
    T=T_o+A_1\frac{\omega^2}{4\left(\sigma-\sigma_M\right)^2+\omega^2},
\end{equation}
where $\sigma_M$ is the optimum noise level and $\omega$ the width of the resonance. As can be seen in Fig.~\ref{experience_theorie} (thin solid lines), this model reproduces very well the resonance of the transmissions for low amounts of noise. However, it does not correctly describe the evolution of the transmission for large values of $\sigma$. It could not be explained by a hypothetic inhomogeneous broadening of the resonant features by increasing noise, neither by a broadening of the discrete level in the Fano model.

The residual part of the transmission at high noise deviations for $L=400\,\mu$m (Fig.~\ref{experience_theorie}B) is very similar to a sigmoid function \cite{LivreAtkins}, which describe the statistic equilibrium between two states. The total evolution of $T$ with $\sigma$ can then be described as the combination of SR theory and classical two level equilibrium theory
\begin{equation}
    T=T_o+A_1 \frac{\omega^2}{4\left(\sigma-\sigma_M\right)^2+\omega^2} + \frac{A_2}{1+\exp\left[\left(\sigma-\sigma_o\right)/\Delta \sigma\right]},
  \end{equation}
Fits are shown in Fig.~\ref{experience_theorie}. A very good agreement between fits and experiments is observed, in both the resonance domain and the decreasing domain. The resonance width $\omega$ for $L=300\,\mu$m and $L=400\,\mu$m are found to be 21.6\,$\mu$m and 9.1\,$\mu$m, respectively. The resonance width characterizes the sensitivity of the SR process with the period $L$ of hole on the screen, and highlights the sensitivity of SPPs generation and propagation to the relation between noise and periodicity. As $\sigma$ increases, the arrays become closer to random hole arrays in which resonant surface waves are absent. The fraction of increased energy coupled by the resonant wave is found to be 0.14$\pm$0.01 compared to random arrays (see Fig.~\ref{experience_theorie}B).
 
In conclusion, the idea that the strong abnormal transmission only originates from the perfect periodic structure of subwavelength hole arrays should be revised. Indeed, extra energy can be coupled through the array by adding a small controlled amount of noise to the position of the holes Energy increase of about 20\% has been reported, 5 fold larger than the relative noise deviation. This new transmission can be modelled with stochastic resonance theory. The evolution of the transmitted energy can be modelled as an equilibrium between a perfect organized and a completely random state, leading to the extra electromagnetic energy coupled at the surface of the screen.

From these experiments two main axes of reflection are drawn. In both visible and near infrared range subwavelength hole arrays experiments, SR effects will have important consequences. Errors on hole position of few nanometres can strongly change coupled energy and may alter experimental reproducibility and this may be a limiting factor to size decreasing in all plasmonic technologies. Furthermore, since small structural modifications have strong effects, stochastic resonance should be a powerful help in detection-related technologies. It may also lead to new interpretations and new applications in three dimensional systems dealing with phonon propagation and photonic crystals technology.
 
\section*{Materials and Methods}

The samples are free-standing 8-$\mu$m-thick nickel-plate arrays of subwavelength squared holes, with $L=300\,\mu$m and $L=400\,\mu$m period, fabricated by electroforming. In all samples, the ratio of the surface of metal over the surface of holes is equal to 3 so that a major part of the transmitted signal is established via the SPPs. Noised arrays are generated so that the minimal distance between any parts of two nearby holes remains larger than 5\,$\mu$m. All arrays have more than one thousand apertures so that the statistic of hole displacement is as close as possible to a perfect Gaussian statistic. The precision over the hole size and periodicity is better than 1\,$\mu$m. For each standard deviation, a difference plate has been used, so a total of 20 plates for $L=300\,\mu$m and 30 plates for $L=400\,\mu$m. The enhanced transmission of the arrays is measured by terahertz time-domain spectroscopy (THz-TDS) \cite{TA8}. Broadband linearly polarized subpicosecond single cycle pulses of terahertz radiation are generated and coherently detected by illuminating photoconductive antennas with two synchronized femtosecond laser pulses. Numerical Fourier transform of the time-domain signals gives access to the characteristic transmission spectrum of the arrays. The transmitted electric field is recorded during 240\,ps, yielding to a 3\,GHz frequency precision after numerical Fourier transform, with $10^4$ signal to noise ratio in $300$\,ms acquisition time. A reference scan is taken with empty aperture. The transmission of the metal array is calculated by the amplitude ratio of the complex spectra of the metal plate and reference scan.

\newpage 

\begin{figure}[tbp]
\begin{center}
\end{center}
\caption{Example of arrays, from a perfect ordered array (left), to a noised array (right).}
\label{array}
\end{figure}

\begin{figure}[tbp]
\begin{center}
\end{center}
\caption{Example among the 30 plates used in the experiment of transmitted energy spectra for $L=$400$\,\mu$m-period plates with noise deviation $\sigma$=0 (black line), $\sigma$=12$\,\mu$m (red line) and $\sigma$=60$\,\mu$m (green line).}
\label{spectre}
\end{figure}

\begin{figure}[tbp]
\begin{center}
\end{center}
\caption{Evolution of the energy transmission with the noise deviation $\sigma$ for $L=$300$\,\mu$m (\textbf{A}) and $L=$400$\,\mu$m (\textbf{B}). The black dots are the experimental data (respectively 20 and 30 plates with different standard deviations). The thick red solid lines are theoretical transmissions. The thin solid black lines are models only taking into account stochastic resonance. Black lines in (\textbf{B}) are other simulations: Bloch wave (dashed), Bethe theory (dotted) and CDWE (dash-dotted). Fitting parameters are for $L=$300$\,\mu$m (\textbf{A}): $\sigma_M=10.3\,\mu$m, $\omega=22.0\,\mu$m, $\sigma_0=37.1\,\mu$m and $\Delta \sigma=7.2\,\mu$m ; for $L=$400$\,\mu$m (\textbf{B}): $\sigma_M=15.3\,\mu$m, $\omega=9.1\,\mu$m, $\sigma_0=42.8\,\mu$m and $\Delta \sigma=4.1\,\mu$m.}
\label{experience_theorie}
\end{figure}

\begin{figure}[tbp]
\begin{center}
\end{center}
\caption{(\textbf{A}) Fano model of a subwavelength hole array with a continuum of states interacting with a discrete resonant state. (\textbf{B}) Fano model including Array Enhanced Stochastic Resonance (AESR).}
\label{schema}
\end{figure}

\newpage
\printfigures

\newpage
\begin{center}
\includegraphics[width=460pt]{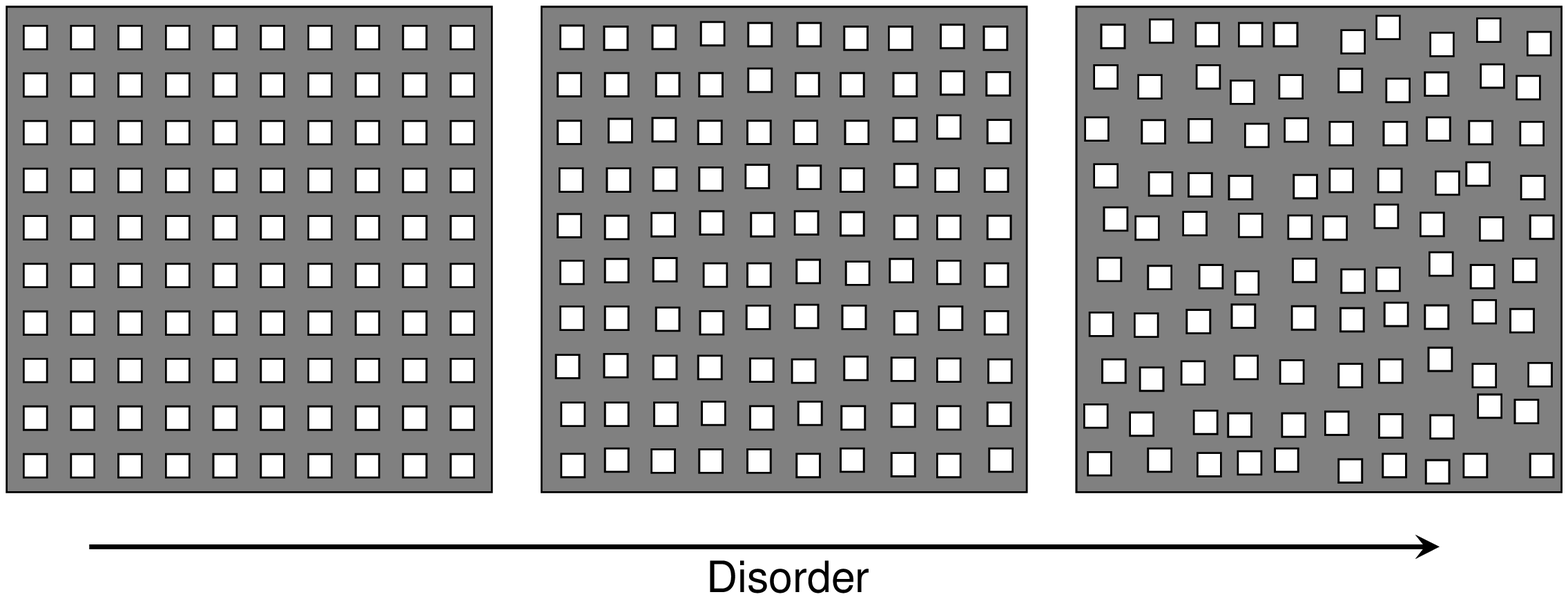}

\vspace{4cm} Masson et al. Figure \ref{array}
\end{center}

\newpage
\begin{center}
\includegraphics[width=460pt]{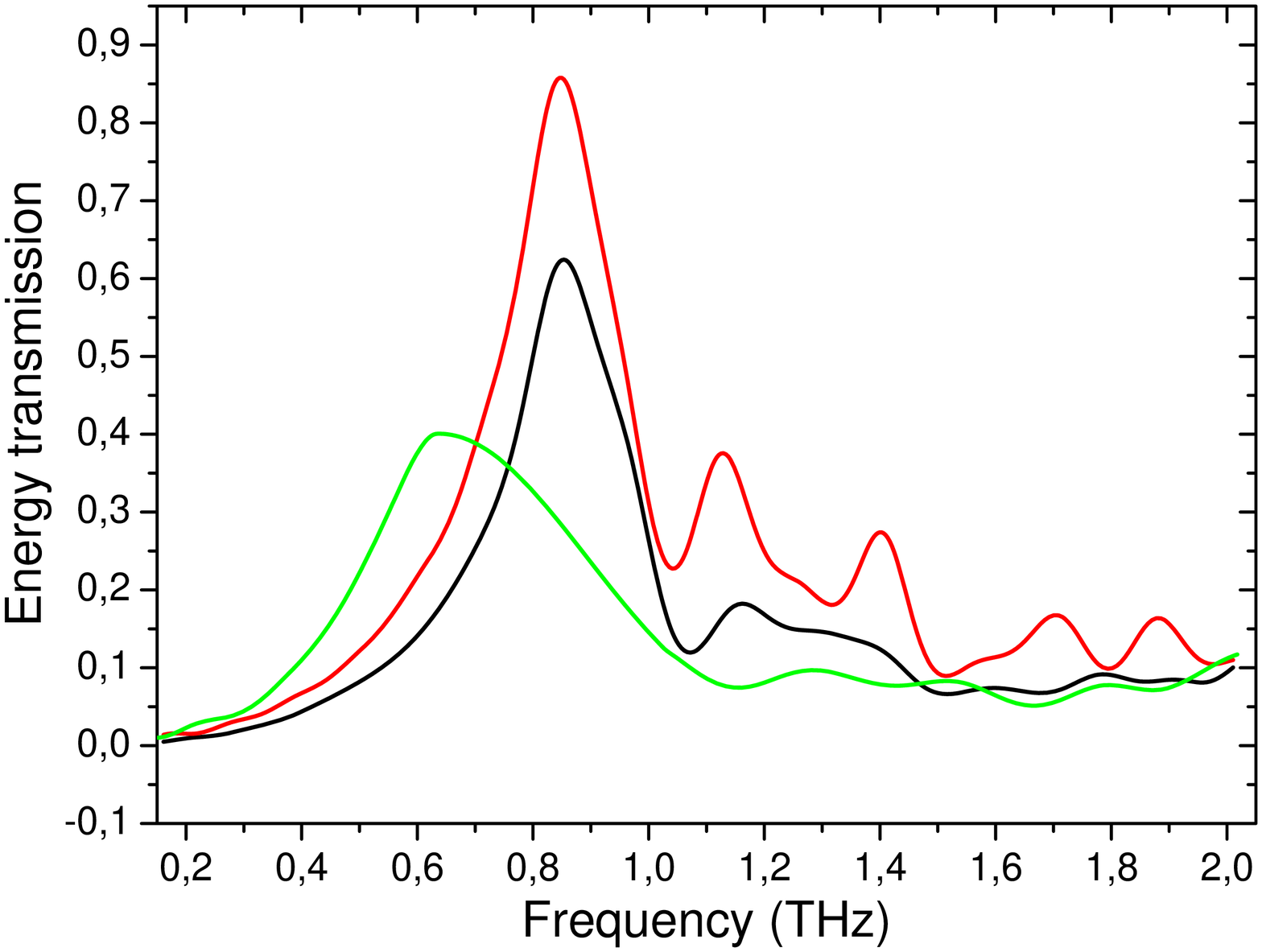}

\vspace{4cm} Masson et al. Figure \ref{spectre}
\end{center}

\newpage
\begin{center}
\includegraphics[width=350pt]{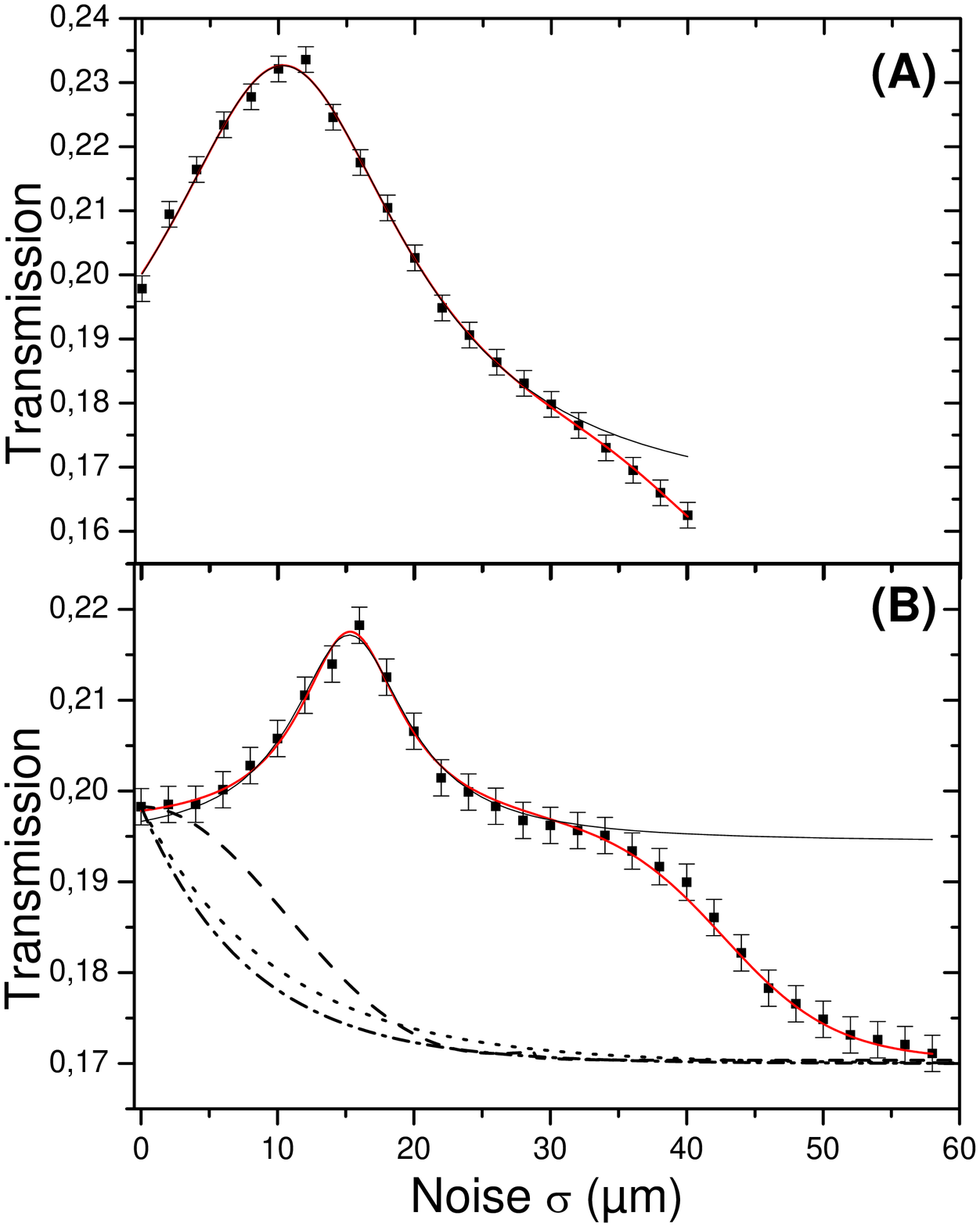}  

\vspace{4cm} Masson et al. Figure \ref{experience_theorie}
\end{center}

\newpage
\begin{center}
\includegraphics[angle=0,width=460pt]{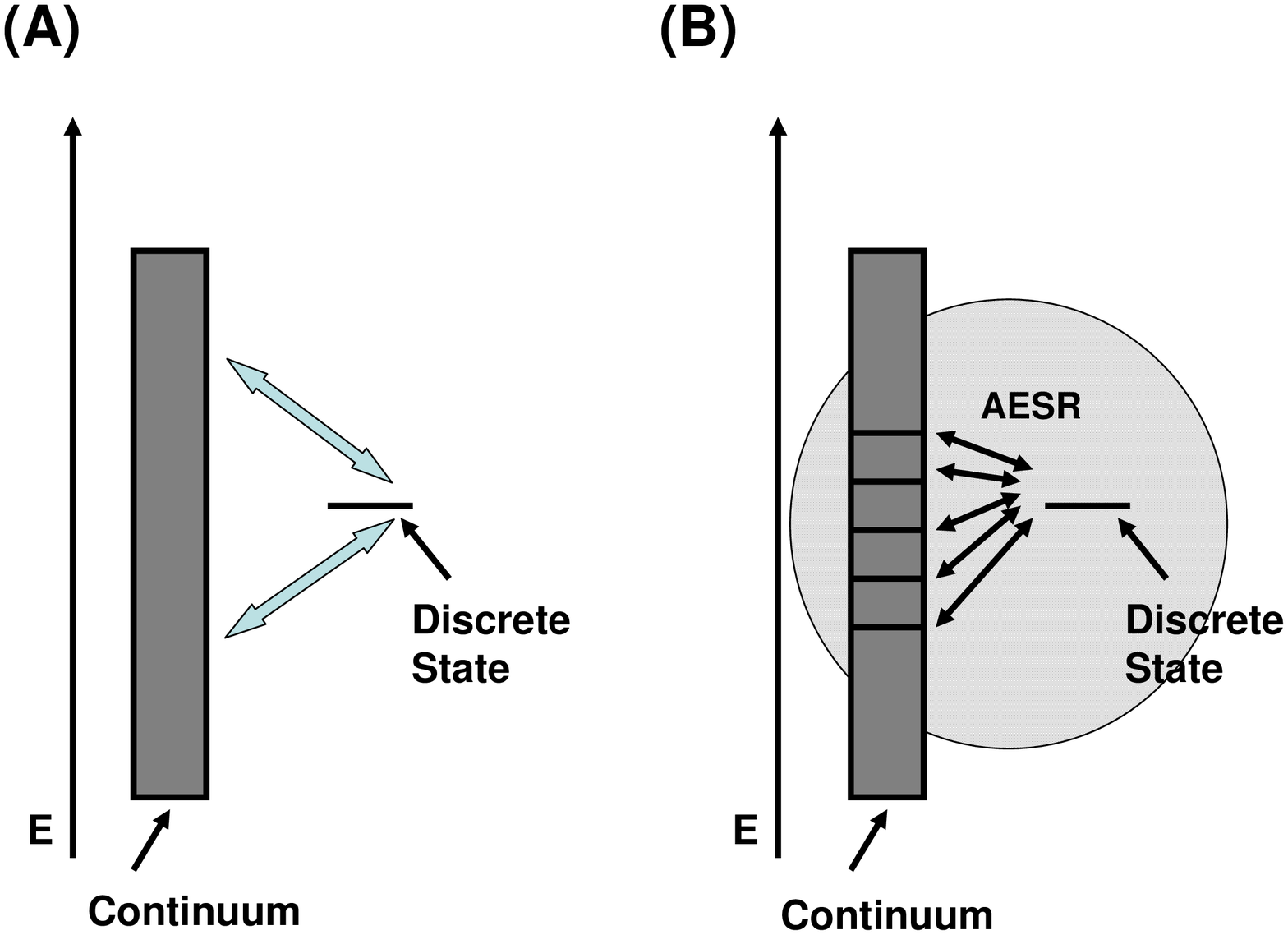}

\vspace{4cm} Masson et al. Figure \ref{schema}
\end{center}

\end{document}